\documentstyle[preprint,eqsecnum,aps,prb]{revtex}
\begin{document}
\draft
\title{Polymorphism and metastability in NbN:\\
Structural predictions from first principles}
\author{Serdar \"{O}\u{g}\"{u}t and Karin M. Rabe}
\address{Department of Physics, Yale University\\
P. O. Box 208120, New Haven, Connecticut, 06520-8120}
\date{10 May 1995}
\maketitle
\begin{abstract}

We use {\em ab initio} pseudopotential total energy
calculations with a plane wave basis set
to investigate the structural energetics of various phases
of polymorphic NbN. Particular attention is given to its
recently discovered superconducting phase with a $T_c$ of 16.4 K,
reported to have the NbO structure type. Results of total
energy calculations show that it is in fact energetically
unfavorable for NbN to form in this cubic structure,
and its predicted theoretical lattice constant is significantly
smaller than the experimental value.
Various approaches to the identification of an alternative
structure are discussed.
In preliminary investigations, we have found
two new structures that are energetically more
favorable than the NbO structure.
\end{abstract}


One of the most important advantages of first principles total
energy and bandstructure calculations as a
tool to understand the physics of materials lies in their
applicability to both real and hypothetical systems.
For example, with
first principles calculations one
can identify structures that are energetically competitive with the
experimentally realized ones, and such efforts can be quite
rewarding if they result in the prediction and identification of new
metastable phases. Combined with the recently developed techniques
for synthesis and preparation of metastable phases, first principles
methods, therefore, prove to be a very useful theoretical
tool to investigate the physics of polymorphic materials.

Recently, a new metastable superconducting phase of NbN
with a $T_c$ of 16.4 K
has been synthesized using the pulsed laser deposition
technique.\cite{Treece1} This phase of NbN,
which is stabilized by heteroepitaxial growth on a MgO (100)
substrate, forms for a narrow range of substrate temperature,
and has been reported to have a primitive
cubic lattice based on the presence of (100) and (300)
peaks in the diffraction pattern that are
forbidden for a face-centered-cubic lattice.
Using X-ray diffraction patterns and oscillation
photographs from this new NbN thin film, it has been
proposed that it crystallizes in the NbO structure
type\cite{Bowman}  with a lattice constant
of 4.442 \AA. NbN has long been known to be a
highly polymorphic material, with five other crystalline
structures already reported in the literature.\cite{Villars}
These include the NbN, AsTi,
NiAs, CW, and the well-known superconducting NaCl ($B1$)
structure types. Such a high degree of polymorphism,
including two superconducting
phases with relatively high $T_c$'s, motivates the
investigation of the structural energetics of this material
and the implications for the occurence of superconductivity.
In this paper, we use first principles total energy
calculations to understand this interesting system, with
particular emphasis on the structure of its recently
discovered superconducting phase.

In our calculations, we used the {\em ab initio} pseudopotential
method with a plane wave basis set and
the conjugate gradients algorithm.\cite{Payne}
The general technical issues that arise in the application
of this method to metallic systems containing transition
metals have been recently discussed in Ref. \onlinecite{Ogut1}.
In this study, for both N and Nb, we constructed optimized
pseudopotentials\cite{Rappe} to perform accurate and
efficient calculations with a plane wave basis set.
For Nb, we used scalar relativistic pseudopotentials, and
treated the $4s$ and $4p$ semicore states as
valence orbitals.
These pseudopotentials were put into separable form with two
projectors\cite{Proj} for each angular momentum with $l=1$ and
$l=2$ components taken as local for N and Nb, respectively.
In order to obtain a 1mRy
kinetic energy convergence, we performed all calculations
at an energy cutoff of 60 Ry. All calculations were performed
within the local density approximation (LDA) with
the exchange-correlation functional of Ceperley and Alder as
parametrized by Perdew and Zunger.\cite{Excorr} {\bf k}-point sets
were constructed using the Monkhorst-Pack scheme\cite{Monkhorst}
with varying choices of $q$, as will be specified below.
To overcome problems associated with finite
{\bf k}-point sampling in metallic systems, we used the
thermal broadening scheme of Gillan\cite{Gillan} with broadening
energies of $\approx$ 0.2 eV. For cubic structures,
the total energies were fitted to the Birch form.
For hexagonal and tetragonal lattices, both the unit cell
volumes and the $c/a$ ratios were varied, and the energies were
fitted to third order polynomials in $c/a$ and to the Birch form
in unit cell volume, as described in
Ref. \onlinecite{Mehl}. After the optimum
volumes and $c/a$ ratios were obtained, the {\bf k}-point grids
were increased to check for convergence. Through the use
of thermal broadening, the convergence as a function of
the number of {\bf k} points was found be very good, as expected,
with structural energy differences differing only a few meV
for larger Monkhorst-Pack grids.

We first performed total energy calculations for the relatively
simple phases of NbN that are observed experimentally, namely the
CW, NiAs, and $B1$ structure types. The CW (space group
$P\bar{6}m2$) and NiAs (space group $P6_3/mmc$) structures
are hexagonal with one and two formula units in the unit cell,
respectively. As shown in Table I, the theoretical lattice parameters
for these structures were found to be in very good agreement with
experiment. Figure 1 shows the results of total energy calculations
for these compounds, and it can be seen that the superconducting
$B1$ phase is actually a metastable phase, as already pointed
out in Ref. \onlinecite{Alekseev}.
The energy difference we find
between the $B1$ and the CW phases is 0.31 eV per formula unit,
while the ground-state energies for the CW and NiAs structures are
almost the same (differing only by 13 meV) within
the calculational accuracy. Next, we performed
total energy calculations for the NbO structure type (space group
$Pm3m$), which has been reported for the
recently discovered superconducting phase of NbN.
This structure can be viewed as a vacancy-ordered $B1$ structure
with the atoms at (0,0,0) and
$(\frac{1}{2},\frac{1}{2},\frac{1}{2})$ missing
[Fig. 2(a)]. As can be seen in Fig. 3, the calculated
energy difference between the NbO and $B1$ structure is 0.92 eV
per formula unit, indicating that if the new phase had the
NbO structure type, it would be extremely high in energy.
Moreover, the theoretical lattice constant 4.207 \AA\ is
significantly smaller
(5.3\%) than the experimental value of 4.442 \AA.
Such a discrepancy between
experiment and theory is suspiciously large.
First of all, the agreement
between our LDA lattice constant for the $B1$ structure and
experiment is very good, with only a 0.5\% underestimate,
indicating that LDA errors are quite small. Second,
although thin films grown epitaxially tend to have slightly
larger lattice constants than their bulk phases, the differences
for a number of NbN$_x$ compounds
are within 1-2\%.\cite{Treece2} Third,
given an experimental lattice constant of 4.39 \AA\ for NbN in the
$B1$ structure, one would expect that with the removal of 25\%\ of
the atoms, the lattice constant for the NbO structure
would be smaller, not larger, although the volume
per formula unit may increase. This expected decrease
in the lattice constant is exactly what is observed
in our previous calculations for the NbO compound.
Our theoretical values for the hypothetical
$B1$ and actual NbO structure types of NbO compound are
4.40 \AA\ and 4.18 \AA,\cite{Ogut2}
respectively, while the experimental value for
NbO is 4.21 \AA.\cite{Bowman} That is, the calculated
reduction in the lattice constant of NbN
as the structure is changed from $B1$ to NbO is very similar
to that of the NbO compound. Therefore, from both
lattice constant values and energetic considerations, our total
energy calculations strongly suggest that the crystal
structure for the recently discovered superconducting
cubic phase of NbN
is not isomorphic to the NbO structure type.

These considerations have led us to search
for other candidate structures for this new phase of
NbN. First, we examined all space groups with a primitive cubic
lattice in the International Tables,\cite{Tables} and considered
all possible crystal structures with three, four, or five formula
units in the primitive cell, since structures with less than three
and greater than five formula units would most likely not result
in a lattice constant of $\approx$ 4.4 \AA.
In addition to the NbO and $B1$ structures, we found
two more possibilities, namely the space group $P23$ with
Nb atoms at the 4(e) positions and N atoms at the 3(c) and 1(a)
positions, and the space group $P4_332$ with Nb atoms at the 4(a)
and N atoms at the 4(b) positions. In the first space group,
the 4(e) positions
$(x,x,x),(x,\bar{x},\bar{x}),(\bar{x},x,\bar{x}),(\bar{x},\bar{x},x)$
are the vertices of a tetrahedron centered at the origin.
For $x=1/4$ this results in a zincblende ($B3$)
structure, which has an fcc lattice. Away from $x=1/4$, although
the lattice becomes simple cubic, our calculations for NbN
show that this distortion
results in a higher energy structure.
For the second
space group $P4_332$, with a lattice constant
of 4.4 \AA, some of the interatomic
distances are much too short (for example, one Nb-N bond length
would have to be 1.55 \AA),
resulting in an energetically very unfavorable structure.
Therefore, there is no plausible
primitive cubic structure with full occupations of the
Wyckoff positions that has a lattice constant close to the
experimentally observed value.

One way to resolve this puzzle is to recognize that
the available experimental data do not absolutely rule out
the possibility of a slightly distorted cubic structure.
If we consider tetragonal structures,
additional vacancy-ordered derivatives of the
$B1$ structure are allowed which
may be energetically more favorable
than NbO and have lattice constants closer to the original.
One such structure has already been discussed in the context of
the NbO compound.\cite{Burdett} It is obtained
from the $B1$ structure by removing the chain of atoms in the [001]
directions, instead of removing the atoms at (0,0,0) and
$(\frac{1}{2},\frac{1}{2},\frac{1}{2})$,
which results in the NbO structure type. This alternative
structure [Fig. 2(b)] also has three formula units in the primitive
cell, however it possesses tetragonal symmetry (space group $P4/mmm$)
because of the preferred direction
along which the atoms have been removed.
Although this structure has the same concentration of vacancies
as NbO, the fact that entire $B1$-type layers are left intact
can be expected to reduce the collapse of the lattice constant
relative to the $B1$ structure.
We minimized the total energy for this $P4/mmm$ structure with
respect to the $c/a$ ratio and volume.
As shown in Fig. 3,
the optimum structure is energetically more favorable than the
NbO structure by an appreciable 0.2 eV per NbN.
Also, the calculated $ab$-plane lattice constant of 4.275 \AA\
for this structure is significantly larger than that of the NbO.
The calculated lattice parameter in the $c$-direction is 4.182 \AA,
so that the unit cell volume is 2.6\%\ larger
than NbO, and the $c/a$ ratio of 0.978 is close to 1. Therefore,
the positions of the Nb atoms, which are most sensitive to
the X-rays, are almost identical to those in the NbO structure.
However, the lattice constants of this structure (particularly
in the $c$-direction) still show
a considerable discrepancy with the experimental observations,
and it is still not competitive in energy with
the $B1$ structure (higher by 0.72 eV per formula unit).
Therefore, although it may be possible to synthesize this phase
under appropriate conditions, we are led to the conclusion that
the observed phase is not a
vacancy-ordered derivative of the $B1$ structure.\cite{Note}

In the search for the correct crystal structure of the new
phase of NbN, it seems that two complementary approaches
should be followed. In the first approach, one can simply
concentrate on structures with potentially lower energies
by identifying the general trends observed in transition metal
nitrides, one of which is the close-packing of the transition
metal atoms with N atoms filling the interstitials.\cite{Wells}
Our calculations along this line found yet another structure
that is energetically more favorable than the NbO structure.
In particular, the hypothetical $B3$ structure of NbN
(where the N atoms are at the tetrahedral interstitials)
was calculated to have a lattice constant of 4.71 \AA\
with a total energy 0.18 eV
per NbN less than the NbO structure type. Although the
appearance of (100) and (300) peaks in the diffraction pattern
rules out this fcc structure as a possible candidate
for the new NbN, again it may be possible to synthesize it
with a suitable substrate and processing conditions.
On the other hand, a theoretical lattice constant of 4.71 \AA\
for the ideal four formula unit cubic cell
makes a vacancy derivative of this structure much more
promising for the experimentally observed new phase of NbN
from the point of view of the lattice constant.

A complementary approach is to concentrate on the
crystallographically allowed structures that are compatible
with the existing experimental observations.
For example, one could consider
all tetragonal space groups with $c/a$ ratios close
to 1 when the Nb-N interatomic distance is taken
as 2.2 \AA.
Another possibility is to allow for fractional occupations
of the Wyckoff positions. For example,
primitive cubic crystal structures with six formula units
and equal fractional occupations of the Wyckoff positions
may have lattice parameters near the experimentally observed
value. This increases the number of candidate structures
considerably. Fractional occupations
are also an interesting possibility
for the vacancy derivatives of the
$B3$ structure.
Starting with an ideal $B3$ structure, removing the Nb
atom at (0,0,0), and assigning a fractional occupancy
of 0.75 for the N atoms at the tetrahedral sites results in
a three-formula-unit cubic structure that could have a
lattice parameter close to 4.44 \AA,
given that the minimum energy lattice
parameter for the ideal $B3$ is 4.71 \AA.
And lastly, there is the possibility that the observed
phase is slightly nonstoichiometric.

In summary, we used the pseudopotential total energy method to
examine the structural energetics of various real and hypothetical
phases of polymorphic NbN. We showed that the NbO structure,
reported for its
recently discovered superconducting phase, is energetically
quite unfavorable, with a theoretical lattice constant
significantly smaller than the experimental value.
Moreover, we argued that no
primitive cubic structure with full occupations of the Wyckoff
positions is a likely candidate for this new phase. In our
general search for alternative structures, we found two
NbN compounds (in $P4/mmm$ and $B3$ structures) that are
energetically more favorable than the NbO structure. Although
these also do not seem to be likely candidates for the new phase,
it may be possible to synthesize them under suitable conditions.
We therefore predict that
there are other metastable phases of the already highly
polymorphic NbN still waiting to be discovered experimentally.
We presented some general approaches that can be followed to
generate a number of candidate structures for the new superconducting
NbN. At this point, more experimental input is needed
to narrow down the theoretical possibilities and to achieve a
definitive determination of the structure.

We would like to thank R. E. Treece, S. C. Erwin,
E. C. Ethridge, L. L. Boyer and
A. R. Kortan for useful discussions and preprints.
This work was supported by NSF Grant No.
DMR-9057442. In addition, K. M. R acknowledges
the support of the Clare Boothe Luce Fund and the Alfred
P. Sloan Foundation.

\begin{figure}
\caption{Total energy versus volume for the CW, NiAs, and $B1$
structures of NbN. The energies are with respect to the minimum energy
corresponding to the LDA lattice parameters of the CW structure.
For the CW and NiAs structures, the plotted energies are for the optimum
$c/a$ ratio of each volume.}
\end{figure}

\begin{figure}
\caption{The crystal structures for the (a) NbO and (b) alternative
$P4/mmm$ structures (mentioned in the text) of NbN. Notice that the
$x=\frac{1}{2}$ plane in the $P4/mmm$ structure is left unchanged
relative to the $B1$ structure.}
\end{figure}

\begin{figure}
\caption{Total energy versus volume for the $B1$, NbO, and $P4/mmm$
structures of NbN. The energies are with respect to the minimum energy
corresponding to the LDA lattice parameters of the CW structure.
For the tetragonal $P4/mmm$ structure, the plotted energies are
for the optimum $c/a$ ratio of each volume.}
\end{figure}

\begin{table}
\caption{The calculated and experimental (in parentheses, taken
from Ref. \protect\onlinecite{Villars}) lattice parameters
for various real and hypothetical structures of NbN and the
number of {\bf k} points used in the calculations. The
numbers in parentheses under the fourth column refer to the total
number of {\bf k} points sampled in the full Brillouin zone.}
\begin{tabular}{lccc}
&\multicolumn{2} {c}{Lattice constants}&\\
Structure&a (\AA)&c (\AA)&Number of {\bf k} points\\
\tableline
CW & 2.928 & 2.848 & 17~(125) \\
& (2.940) & (2.790) & \\
NiAs & 2.953 & 5.491 & 10~(75) \\
& (2.968) & (5.549) & \\
$B1$ & 4.370 && 10~(256)\\
& (4.391) && \\
NbO & 4.207 && 4~(64) \\
$P4/mmm$ & 4.275 & 4.182 & 6~(64) \\
$B3$ & 4.712 && 10~(256)\\
\end{tabular}
\end{table}


\begin{references}
\bibitem{Note} There is one other tetragonal vacancy-ordered derivative
of the $B1$ structure,
obtained by removing the atoms at
(0,0,$\frac{1}{2}$) and ($\frac{1}{2}$,$\frac{1}{2}$,0). However,
in this structure none of the $B1$-type layers parallel to the
primitive lattice vectors are kept intact, and our calculations
indicate that the lattice constants are smaller than
those of the tetragonal structure discussed in the text.
\end{references}
\end{document}